\documentclass[
  aps,
  prl,
  reprint,
  superscriptaddress,
  floatfix,
  amsmath,amssymb,
  nofootinbib
]{revtex4-2}

\usepackage[utf8]{inputenc}
\usepackage[T1]{fontenc}
\usepackage{graphicx}
\usepackage{siunitx}
\usepackage{physics}
\usepackage{bm}
\usepackage{url}

\usepackage{verbatim}

\usepackage[normalem]{ulem} 
\usepackage{xcolor}

\newif\ifshowchanges
\showchangesfalse

\newcommand{\change}[1]{\ifshowchanges\textcolor{red}{#1}\else#1\fi}

\newcommand{\del}[1]{\ifshowchanges\textcolor{black}{\sout{#1}}\fi}

\begin{document}

\title{Effective Atom Theory: Gradient-Driven Materials Design}

\author{J.~Tahmassebpur}
\affiliation{Department of Applied and Engineering Physics, Cornell University}
\author{C. M\'endez}
\affiliation{Department of Applied and Engineering Physics, Cornell University}
\author{B.~Li}
\affiliation{Department of Physics, Cornell University}
\affiliation{JILA and Department of Physics, University of Colorado, Boulder}
\author{B.~Barron}
\affiliation{Department of Physics, Cornell University}
\affiliation{Max Planck Institute for Demographic Research, Germany}
\author{H.~Abruña}
\affiliation{Department of Chemistry and Chemical Biology, Cornell University}
\author{P.~Frazier}
\affiliation{School of Operations Research and Information Engineering, Cornell University}
\author{T.~A.~Arias}
\affiliation{Department of Physics, Cornell University}

\date{\today}

\begin{abstract}
We introduce Effective Atom Theory (EAT), which embeds discrete atomic identities into a continuous differentiable space.  During optimization, sites pass through temporary effective mixtures, then are driven to fully discrete, physically realizable atoms. This enables gradient-based optimization in DFT and MLIPs, yielding orders-of-magnitude acceleration over discrete methods. EAT recovers the exact ground state of NaCl in a 128 atom cell, reconstructs the Cu--Pd phase diagram, optimizes high-entropy oxides for oxygen evolution, and discovers bonding topologies as demonstrated for SrTiO$_3$.
\end{abstract}

\maketitle

\textit{Introduction}---The growing demand for sustainable energy, advanced electronics, and lightweight structural components drives an urgent search for materials with tailored properties~\cite{Dresselhaus2001, Geim2007, Miracle2017}. Density-functional theory (DFT) reliably predicts total energies, forces, and electronic properties~\cite{RevModPhys.64.1045}, but its cost makes exhaustive exploration of combinatorial structure-composition space impractical~\cite{pedersen2021BO}. Materials optimization involves two intertwined challenges: optimizing chemical ordering within a fixed bonding topology, and optimizing bonding topology itself. While the bonding-topology problem is well recognized, even fixed-topology optimization faces a combinatorial explosion central to materials families---including perovskites, rock salts, and high-entropy materials---where there are astronomically many elemental arrangements even when the crystal structure is unchanged.

To accelerate discovery, DFT has been coupled with evolutionary/genetic algorithms~\cite{Johannesson2002Evolutionary,Froemming2009Genetic,jennings2019genetic}, Bayesian optimization~\cite{chen2024adaptive, lookman2019active, pedersen2021BO, herbol2018efficient, Chitturi2024, tran2018active, Zhong2020, xu2024}, and machine-learning (ML) surrogates trained on large DFT datasets~\cite{lan2023adsorbml, Janet2020, liu2020, Zhang2022, oc20, oc22}. Even for fixed topology, direct DFT-driven genetic algorithms can require $>10^3$ \emph{ab initio} calculations~\cite{bauer2022GA}. Bayesian optimization (BO) reduces this cost by fitting a probabilistic surrogate and selecting queries through an acquisition function~\cite{frazier2018tutorialbayesianoptimization}, with applications to catalysts, perovskites, and alloy design~\cite{tran2018active,herbol2018efficient,pedersen2021BO,lookman2019active,chen2024adaptive}; yet combinatorial BO can still require hundreds of DFT evaluations in spaces far smaller than those considered below~\cite{park2025active}. Large datasets such as OC20 and OC22 enable graph neural networks yielding $10^{3}$--$10^{4}\times$ speed-ups in adsorbate--surface searches~\cite{oc20,oc22,lan2023adsorbml}, but require millions of \textit{ab initio} calculations, and even the resulting accurate machine-learning interatomic potentials (MLIPs) do not remove topological optimization or fixed-topology combinatorial search problems.

One way to relieve the discrete combinatorial search problem is to let atoms or species transform continuously in an enlarged effective space. Such mixing is not new: in the virtual crystal approximation (VCA)~\cite{eckhardt2014VCA,Bellaiche2000VCA}, all sites share a globally averaged ``virtual atom,'' yielding a spatially averaged crystal rather than a physical ordering of discrete elemental species. Alchemical MLIPs likewise interpolate between species~\cite{nam2025differentiableAlchemical}; however, like VCA, they work within an averaged crystal approximation. Gaussian-process approaches have been devised to interpolate between species during structure optimization~\cite{kaappa2021interpolation} and, more recently, have been applied to the topology optimization problem~\cite{Larsen2025BarrierCircumvention}. These Gaussian-process methods search for materials with discrete elemental species, but rely on specialized surrogates that do not match the underlying energy function at all discrete realizations and recover discreteness by nearest-species rounding, which can lead to non-optimal solutions.

\del{To address these challenges, we introduce Effective Atom Theory (EAT), which replaces discrete chemical-ordering and bonding topology search with a smooth, continuous optimization directly within both DFT and machine learning interatomic potential (MLIP) calculations.}
\change{To address the combinatorial challenge of chemical ordering optimization, we introduce Effective Atom Theory (EAT), which embeds discrete elemental identities in a continuous, differentiable space, enabling gradient-driven optimization directly within DFT and MLIP energy models. We additionally demonstrate a preliminary extension of this framework to bonding-topology optimization.} In EAT, each atom or ionic core $I$ is assigned mixing coefficients $x_{I\alpha}$ over elements $\alpha$, subject to simplex constraints, $x_{I\alpha}\geq 0, \ \sum_\alpha x_{I\alpha}=1$, making the total energy $E[\{x_{I\alpha}\}]$ differentiable in these variables. This converts materials optimization from discrete sampling to gradient-based optimization, solvable with standard quasi-Newton methods that converge in orders of magnitude fewer iterations than typical GA or BO runs. Crucially, EAT introduces no surrogate: at every discrete embedded point it recovers the corresponding physical DFT or MLIP material exactly. Moreover, we introduce ``syntropization'' to continuously drive each $x_{I\alpha}$ toward $0$ or $1$ while recomputing material properties, yielding optimal physically realizable systems with no probabilistic atomic mixtures. Below, we demonstrate EAT in a wide range of applications from chemical-order optimization, phase-diagram construction, and catalytic-site optimization, and then show how to generalize EAT to bonding-topology optimization.

\textit{New Method}---We consider a unit cell with atomic sites occupied by either real atoms or \emph{effective} atoms. Each effective atom at site $I$ is a continuous mixture over species $\alpha$ with coefficients $x_{I\alpha}$. In DFT, effective atoms are constructed analogously to the VCA~\cite{Bellaiche2000VCA} (as in VASP) via linear combinations of elemental pseudopotentials.  For MLIPs, they are constructed in our FAIRChem\cite{wood2025uma} implementation by linearly mixing elemental node embeddings and source/target edge embeddings. In both cases total energies $E_n$ are smooth in $x_{I\alpha}$, and their gradients are available with negligible overhead: via Hellmann--Feynman in DFT after ground-state convergence (no extra self-consistent-field (SCF) cycles needed), and via backpropagation in FairChem. (See End Matter for details.) We can thus use standard gradient‐based algorithms to optimize any property $f(\{E_n\})$ built from total energies $E_n$ of any collection of configurations $n$ of the involved atoms.  To maintain physically reasonable mixtures, we restrict each $x_{I\alpha}$ to the composition simplex 
\begin{equation}
    x_{I\alpha}\ge0, \quad \sum_\alpha x_{I\alpha}=1.
    \label{eq:simplex}
\end{equation}
We then handle the resulting optimization problem efficiently using a projected quasi‐Newton method \cite{Projected_quasi_newton} to find the nearest local optimum. For clarity, the optimization begins at some initial $x_{I\alpha}$ with a full relaxation of all of the internal degrees of freedom needed to evaluate $f(\{E_n\})$---including relaxation of whichever atomic locations and lattice vectors are to be optimized and, in the DFT case, of the electronic coordinates. This is followed by computing the partial derivatives of the objective function with respect to $x_{I\alpha}$, and then using these gradients to update $x_{I\alpha}$ via a quasi-Newton step. At the updated $x_{I\alpha}$, we repeat this procedure.

\begin{figure}[tb]
\parbox{1.1in}{\includegraphics[width=1in]{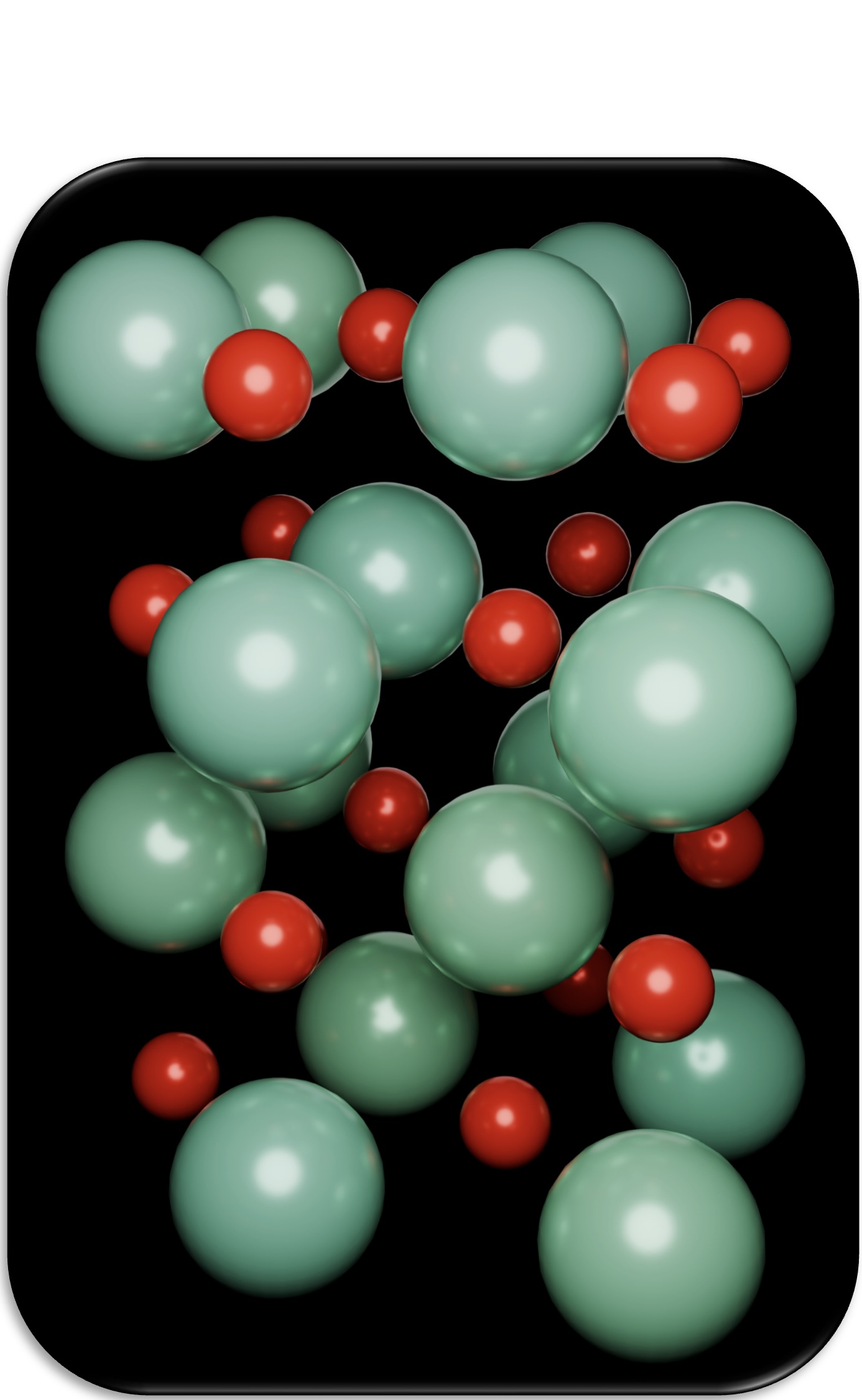}\\(a)}\parbox{1.1in}{\includegraphics[width=1in]{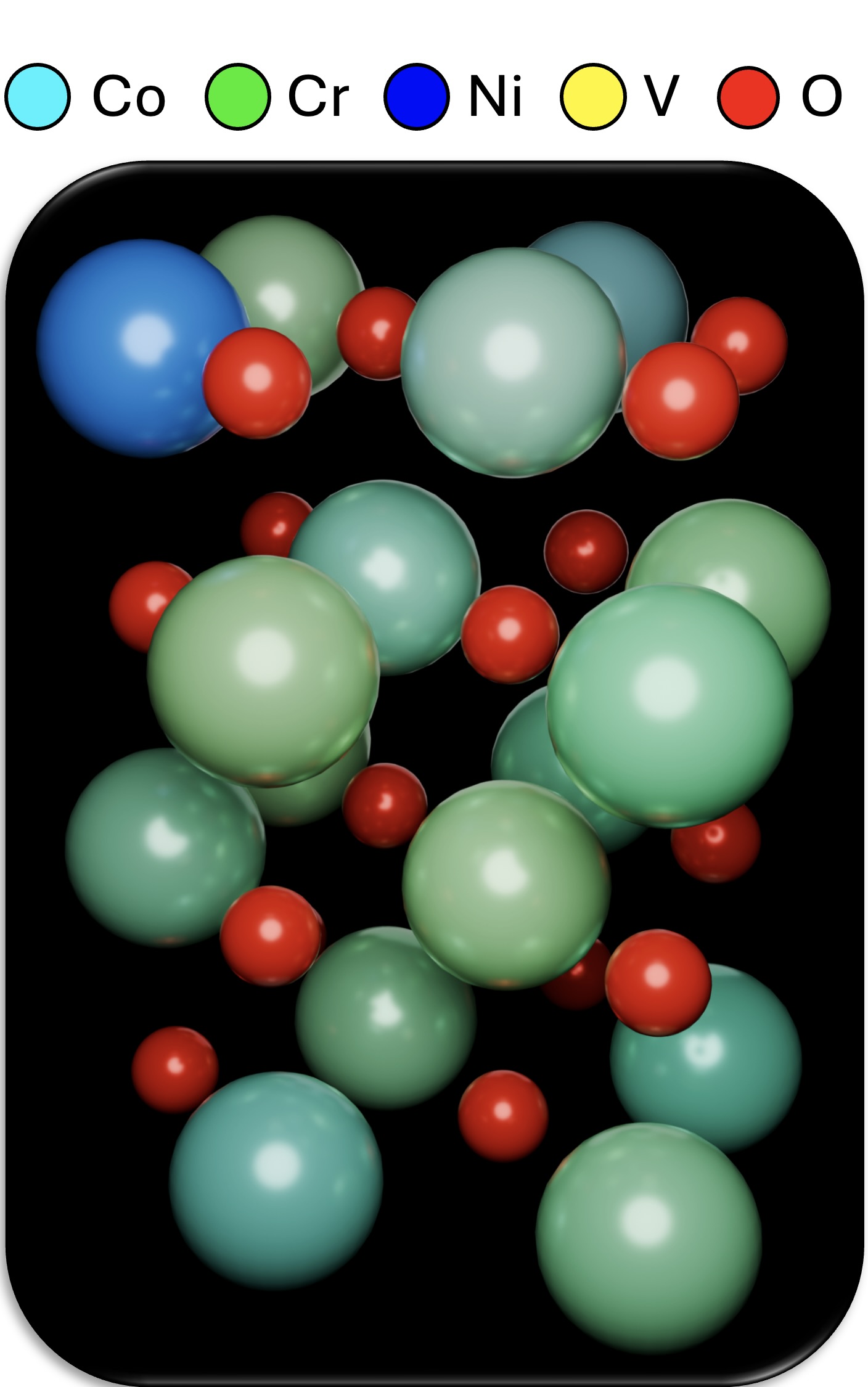}\\(b)}\parbox{1.1in}{\includegraphics[width=1in]{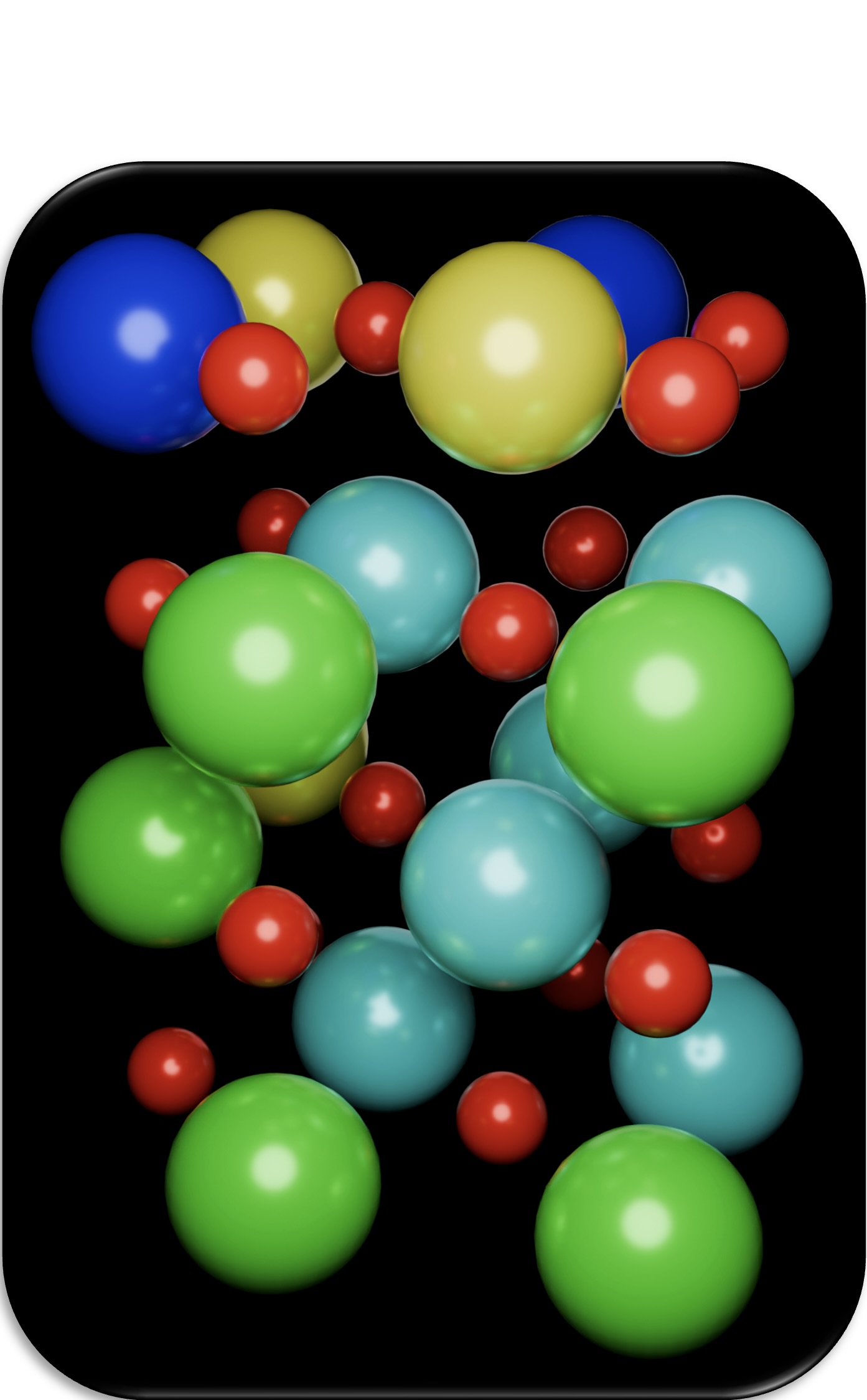}\\(c)}
  \caption{EAT optimization of a unit cell: (a) initial random configuration, (b) optimized effective-atom configuration, and (c) final syntropized configuration of true elemental species. Spheres mark atomic positions, with oxygen shown as smaller red spheres for clarity. Real elements are color-coded as Co (cyan), Cr (green), Ni (blue), and V (yellow). Effective atoms in (a,b) are shaded using RGB mixtures of these colors; for instance, the atom at the upper left of (b) combines 72\% Ni (blue) and 28\% Co (cyan).}
  \label{fig:syntropization}
\end{figure}

The above procedure will generally yield an optimum in which the chemical configuration $x_{I\alpha}$ still contains effective (fractional) atoms. This is of limited practical value, as materials composed of fractional atoms do not exist in nature. To address this, one must introduce a procedure that converts the effective atoms into true elemental species while retaining the optimized target properties. We therefore introduce a penalty that drives optimization away from effective atoms via a ``syntropization'' term that suppresses compositional entropy in the stoichiometric weights $x_{I\alpha}$. The term ``syntropization'' draws on the historical notion of syntropy~\cite{vyatkin2019synergetic}---commonly interpreted as the opposite tendency of entropy---and here signifies the recovery of discrete atomic identities from their effective (fractional) representations. Denoting the penalty by $S(\{x_{I\alpha}\})$, we minimize (note the positive sign)
\begin{equation}
  \mathcal L(\{E_n, x_{I\alpha}\}) \;=\; f(\{E_n\}) \;+\;\lambda\,S(\{x_{I\alpha}\}),
  \label{eq:loss}
\end{equation}
where $\lambda$ tunes the strength of the drive toward elemental stoichiometries.  
One natural choice for the penalty $S(x_{I\alpha})$ is the Tsallis q-entropy ($q\geq1$), which we employ below with $q=2$ to avoid singular derivatives at the simplex vertices,
\begin{equation}
    S(q,\{x_{I\alpha}\}) = \frac{1}{q-1}\sum_{I}\left(1 - \sum_\alpha (x_{I\alpha})^q\right).
    \label{eq:tsallis}
\end{equation}
We then proceed by ramping up the penalty strength $\lambda$ while continually reoptimizing $L(\{E_n, x_{I\alpha}\})$ (see \textit{End Matter} for details). Fig.~\ref{fig:syntropization} illustrates the three key stages of the EAT procedure: (a) initial randomly mixed species for each atom, (b) optimal effective atoms, (c) final optimal elemental atoms resulting from syntropization. 

\textit{Application to finding ground-state chemical ordering}---As an initial demonstration, we use EAT to search for the ground-state chemical ordering of a 128-atom rocksalt NaCl supercell and compare against a leading alternative, discrete Bayesian optimization (BO). Since Na and Cl have differing valence orbitals, we avoid using the DFT implementation of EAT for this application (see the End Matter for an explanation) and instead employ our version of the MLIP from FAIRChem~\cite{wood2025uma}. Fixing the ions and lattice during the optimizations, we minimize the formation energy subject to a composition penalty ensuring 50/50 Na and Cl. Exhaustive enumeration is impossible at this scale: at fixed composition the probability of sampling the exact ground state at random is \del{$\mathcal{O}(10^{-37})$}\change{1 in $10^{37}$}.

\begin{figure}[!tb]
    \centering
    \includegraphics[width=0.8\linewidth]{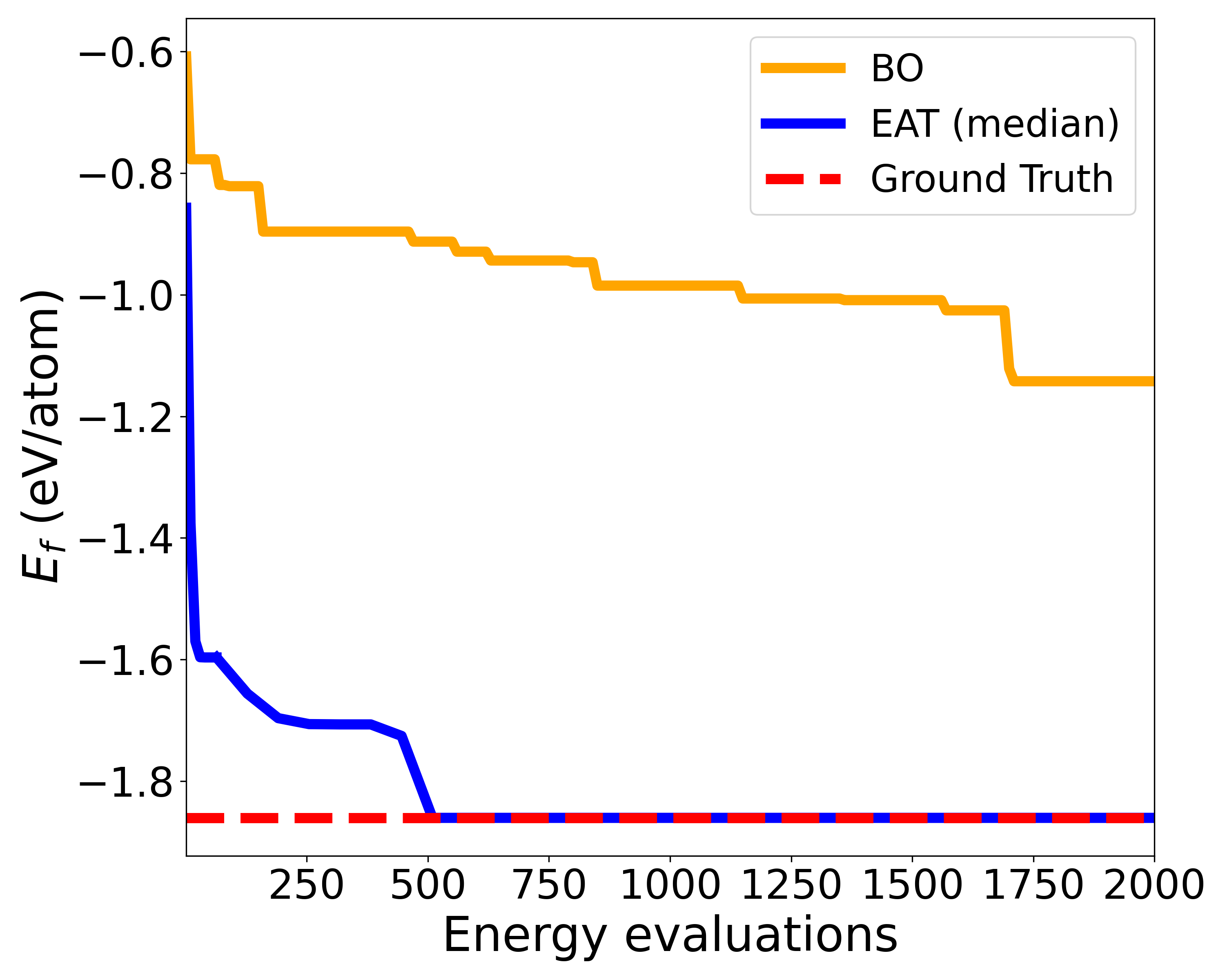}
    \caption{
    EAT vs.\ discrete Bayesian optimization (BO) for the ground-state chemical ordering of a 128-atom rocksalt NaCl supercell at fixed 50/50 composition (FAIRChem energies).
    Blue: median of the set of EAT runs plotted versus effective evaluations (see text and End Matter).
    Orange: BO best-so-far energy.
    Red dashed: ground-truth formation energy.
    }
    \label{fig:NaCl}
\end{figure}

We perform 100 independent EAT runs, each requiring on average $L\approx 60$ iterations to fully syntropize. We compare this against a standard implementation of discrete Bayesian optimization (BO) \cite{balandat2020botorch} (see End Matter for details). Figure~\ref{fig:NaCl} shows that, remarkably, EAT recovers the exact ground-state ordering from among the $\sim 10^{37}$ possibilities in a median of 500 energy evaluations. On the contrary, even after 2000 iterations, beyond which the BO steps themselves become computationally expensive, we find that BO makes relatively little progress.

\textit{Application to phase diagram construction}---With the capabilities of effective atom theory (EAT) demonstrated in the previous application, here we show how EAT can be used to construct the convex hull of the binary Cu--Pd system. For this demonstration, we use both FAIRChem and DFT because the valence orbitals of Cu and Pd are compatible.

For reference, we note that Materials Project \cite{materials_project} reports three stable phases on the Cu--Pd convex hull, at compositions
$\mathrm{Cu_{0.25}Pd_{0.75}}$,
$\mathrm{Cu_{0.5}Pd_{0.5}}$, and
$\mathrm{Cu_{0.75}Pd_{0.25}}$.
All three may be approximately represented by FCC unit cells with lattice constant linearly interpolated between pure FCC Cu and pure FCC Pd (See End Matter). For our construction, 
we discretize the binary phase diagram in $1/8$ composition increments, at each composition we use EAT to search directly for the minimum-formation-energy, fully syntropized chemical ordering. \change{The composition penalty is soft and hard in the FAIRChem and DFT runs, respectively.}

\begin{figure}[!tb]
\parbox{1.7in}{\includegraphics[width=1.7in]{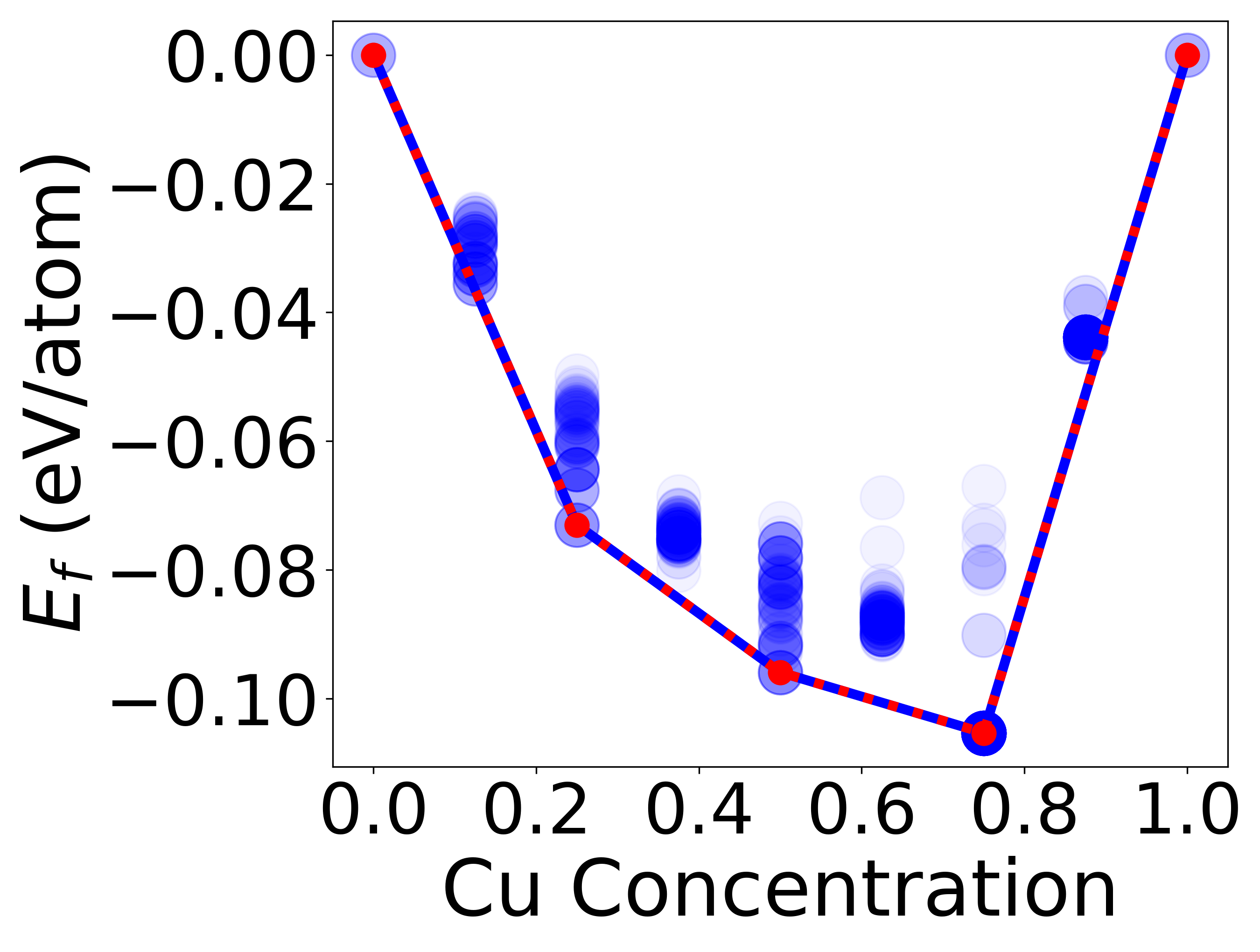}\\(a)}\parbox{1.7in}{\includegraphics[width=1.7in]{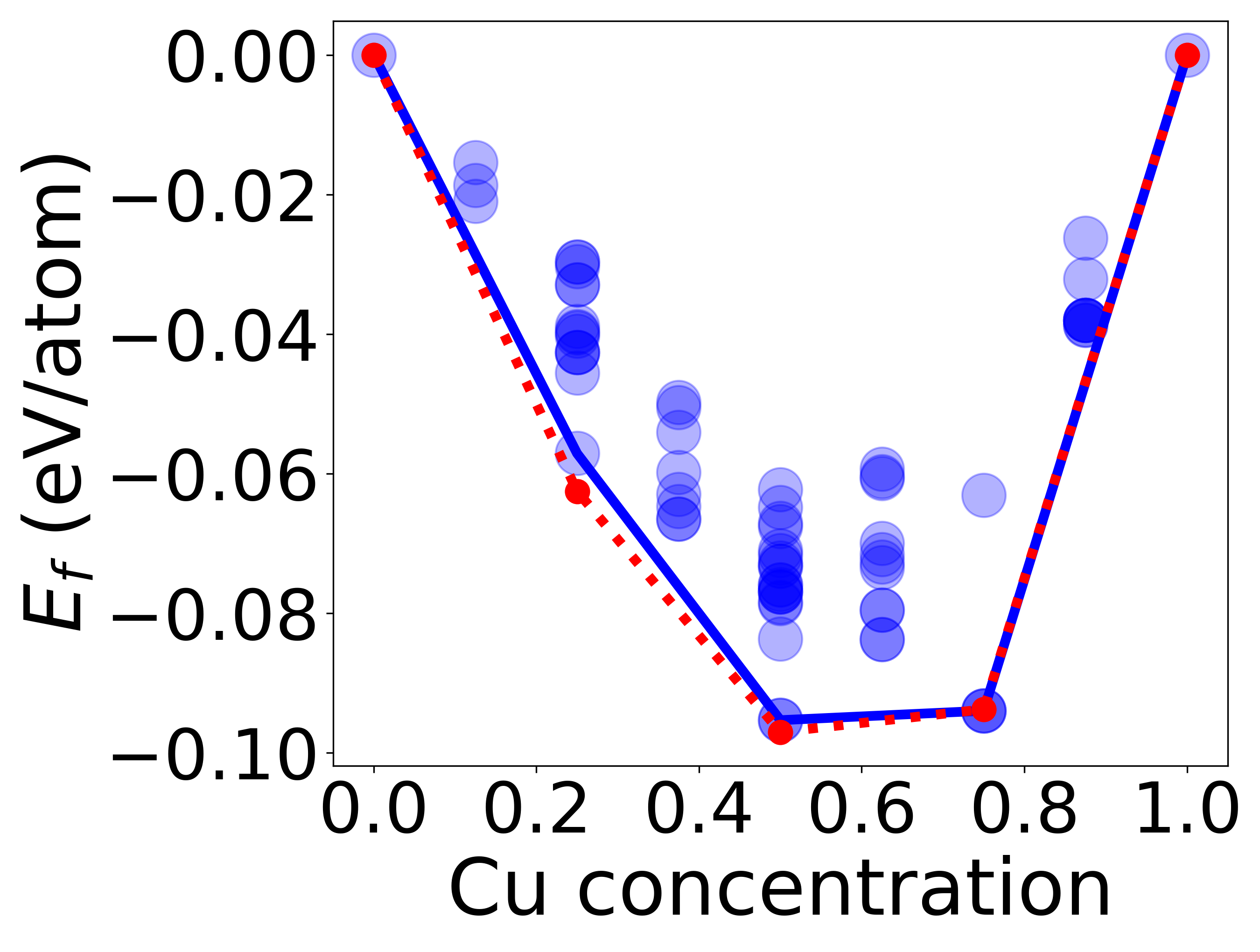}\\(b)}
  \caption{Construction of the Cu--Pd formation-energy convex hull using EAT. (a) FAIRChem-based EAT results in a 32-atom FCC supercell compared with the FAIRChem-calculated ground truth reference hull (red). (b) DFT-based EAT results in a 16-atom FCC supercell compared with the DFT-calculated ground truth reference hull (red). In both panels, blue points show independent EAT optimizations after syntropization, and the blue line is their lower convex envelope.}
  \label{fig:hull}
\end{figure}

We first perform the convex hull construction in a 32-atom $2\times 2\times 2$ conventional FCC supercell using FAIRChem. Despite the explosion of configuration space ($\mathcal{O}(10^8)$ configurations), Fig.~\ref{fig:hull}(a) shows that EAT identifies the ground truth hull exactly, using at most $\sim 10^4$ total energy evaluations per composition. This is roughly $10^4$ times faster than exhaustive enumeration.

Figure~\ref{fig:hull}(b) shows the results of our DFT calculations. To construct the convex hull, we explore configurations in a $1\times 2\times 2$ 16-atom conventional FCC supercell, with the ions and lattice held fixed throughout the optimization. We use a smaller cell here for computational efficiency since each EAT run requires $\sim 40$--$50$ DFT energy evaluations. As the figure shows, across compositions, EAT finds low-energy orderings that either lie exactly on or within 0.01 eV/atom of the hull. In particular, at $\mathrm{Cu_{0.75}Pd_{0.25}}$, EAT recovers the exact ground-state ordering within two independent runs ($\sim 100$ alloy evaluations total), compared with $1820$ discrete orderings in this cell at this composition, implying a speedup of an order of magnitude over exhaustive enumeration. At $\mathrm{Cu_{0.5}Pd_{0.5}}$, EAT likewise reaches within 0.01 eV/atom of the hull in about $200$ alloy evaluations, compared with $12870$ possible orderings. Finally, at $\mathrm{Cu_{0.25}Pd_{0.75}}$, EAT reaches within $0.01$ eV/atom from the hull after about $450$ energy evaluations, compared with $1820$ discrete orderings. We also note that, for this 16-atom cell, FAIRChem finds the exact ground-state hull in roughly 250 energy evaluations per composition, a cost comparable to DFT. This suggests that, for the larger cell, EAT with DFT would also perform comparably to the FAIRChem results.

\textit{Application to OER catalyst design}---To illustrate the use of EAT’s density-functional theory capabilities in a practical application of current interest, and to show that it extends beyond ground-state structure prediction to the optimization of any property governed by energy differences, we next optimize high‐entropy rock‐salt oxides for the alkaline oxygen evolution reaction (OER) within DFT. A good OER catalyst minimizes the overpotential
\begin{align}
  \eta_{\rm OER}
  &= \frac{1}{e}\,
     \max\bigl(\Delta G, 3.2 \ \mathrm{eV} - \Delta G\bigr) - 1.23 \ \mathrm{V},
  \label{eq:overpot}
\end{align}
where $\Delta G = G_{\mathrm{O}^*} - G_{\mathrm{HO}^*}$ is the difference in the adsorption free‐energies of O and OH~\cite{Man_Norskov_Universality_OER}. The optimum occurs when $\Delta G =1.6$~eV. In terms of DFT-computed energies,
\begin{equation}
  \Delta G = E(\mathrm{O}^*) - E(\mathrm{HO}^*) + \tfrac12 E(\mathrm{H_2}) - 0.36\,\text{eV},
  \label{eq:dG}
\end{equation}
with $0.36$~eV being an entropic correction~\cite{Man_Norskov_Universality_OER}.

Despite its single‐descriptor simplicity, the so‐called OER ``volcano'' relation (Eq.~\ref{eq:overpot}) has proven remarkably accurate at capturing trends in oxide overpotentials across a wide range of materials~\cite{Norskov_OER_Experimental_Volcano}. Nonetheless, rock‐salt NiO---whose descriptor value places it tantalizingly close to the volcano’s peak--- actually underperforms in experiment~\cite{Norskov_OER_Experimental_Volcano}. This discrepancy likely stems from factors outside the descriptor itself: NiO’s poor electronic conductivity~\cite{zhang_NiO_poor_conductivity} and its tendency to undergo phase transitions under OER conditions~\cite{hales_NiO_phase_transitions} undermine its catalytic activity.

To overcome NiO’s limitations, we investigate a Co–Cr–Ni–V high-entropy rock-salt oxide: Co doping enhances electronic conductivity~\cite{vazhayil_Co_doped_NiO}, while V and Cr improve OER activity~\cite{jiang_V_Ni_good_OER,lee_Cr_Ni_good_OER}. Figure~\ref{fig:syntropization} shows the (100) facet~\cite{stable_facet_100} rock-salt unit cell in which the metal-site elemental composition is varied; the cell contains 16 metal and 16 oxygen atoms (excluding adsorbates) arranged in four layers, and the lattice constant is set to the average of CoO, CrO, NiO, and VO.

To proceed, we encode the alkaline OER‐volcano target as a least‐squares loss,
\begin{equation}
  f\bigl(E(\mathrm{O}^*),E(\mathrm{HO}^*)\bigr)
  = \tfrac12\bigl(\Delta G - 1.6\,\mathrm{eV}\bigr)^2,
  \label{eq:ols}
\end{equation}
where $\Delta G$ is defined in Eq.~\ref{eq:dG}, and then minimize Eq.~\ref{eq:ols} subject to the simplex constraints (Eq.~\ref{eq:simplex}). All atoms are initially held at fixed locations except the adsorbate, which is relaxed only in the first iteration, before any updates of the stoichiometric weights $x_{I\alpha}$. We also fix the lattice throughout the optimization. For the initial search, we disable spin polarization and Hubbard–$U$ to preserve a smooth, convex DFT energy surface---spin effects introduce multiple self‐consistent solutions and abrupt energy discontinuities. (We eventually include ionic relaxation, spin and Hubbard-$U$ below, after the initial EAT search.) Finally, we perform $15$ independent optimizations to increase the likelihood of reaching the global optimum.

\begin{figure}[tb]
  \centering
  \parbox{1.7in}{\includegraphics[width=1.7in]{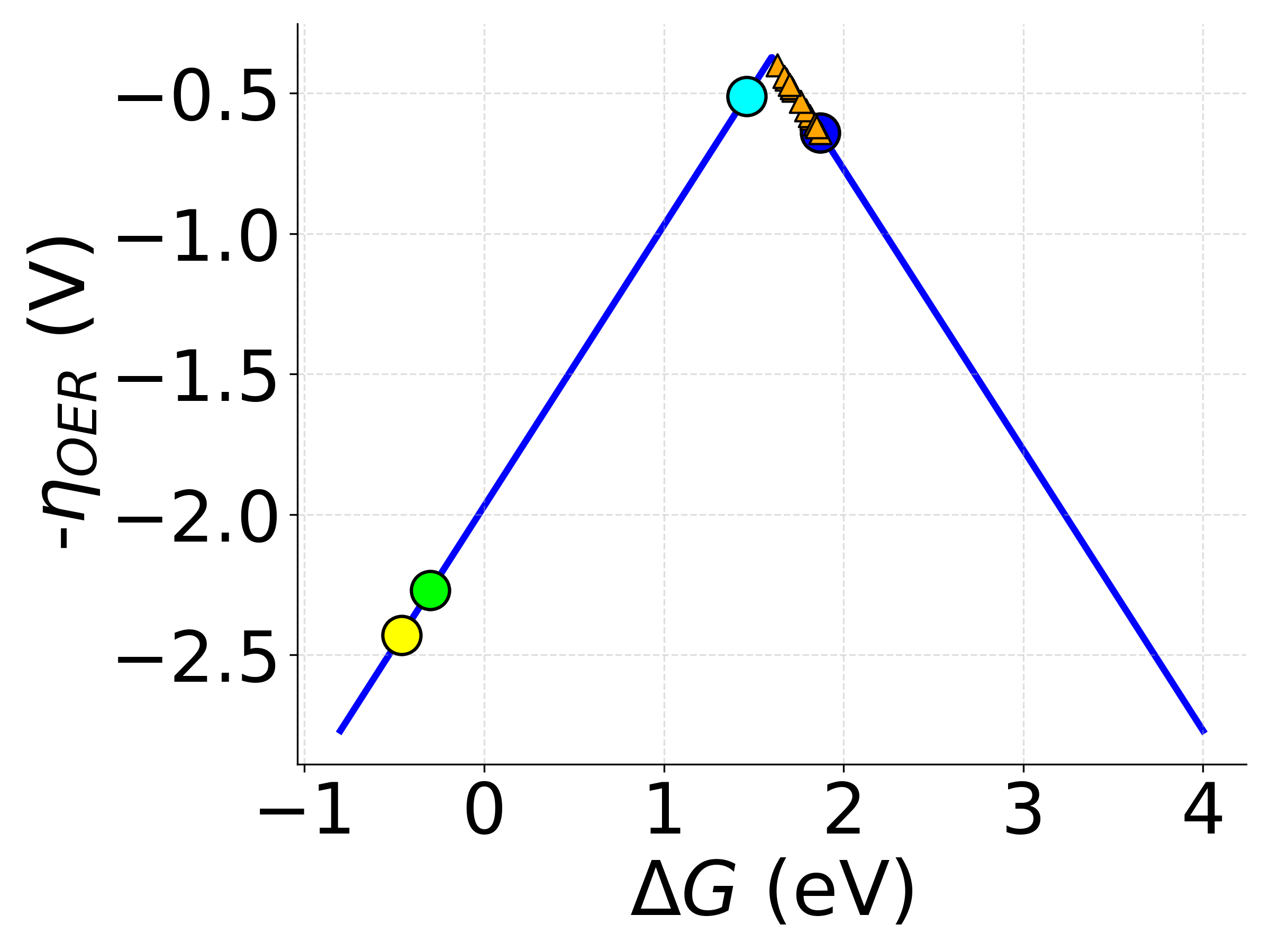}\\(a)}\hfill
  \parbox{1.7in}{\includegraphics[width=1.7in]{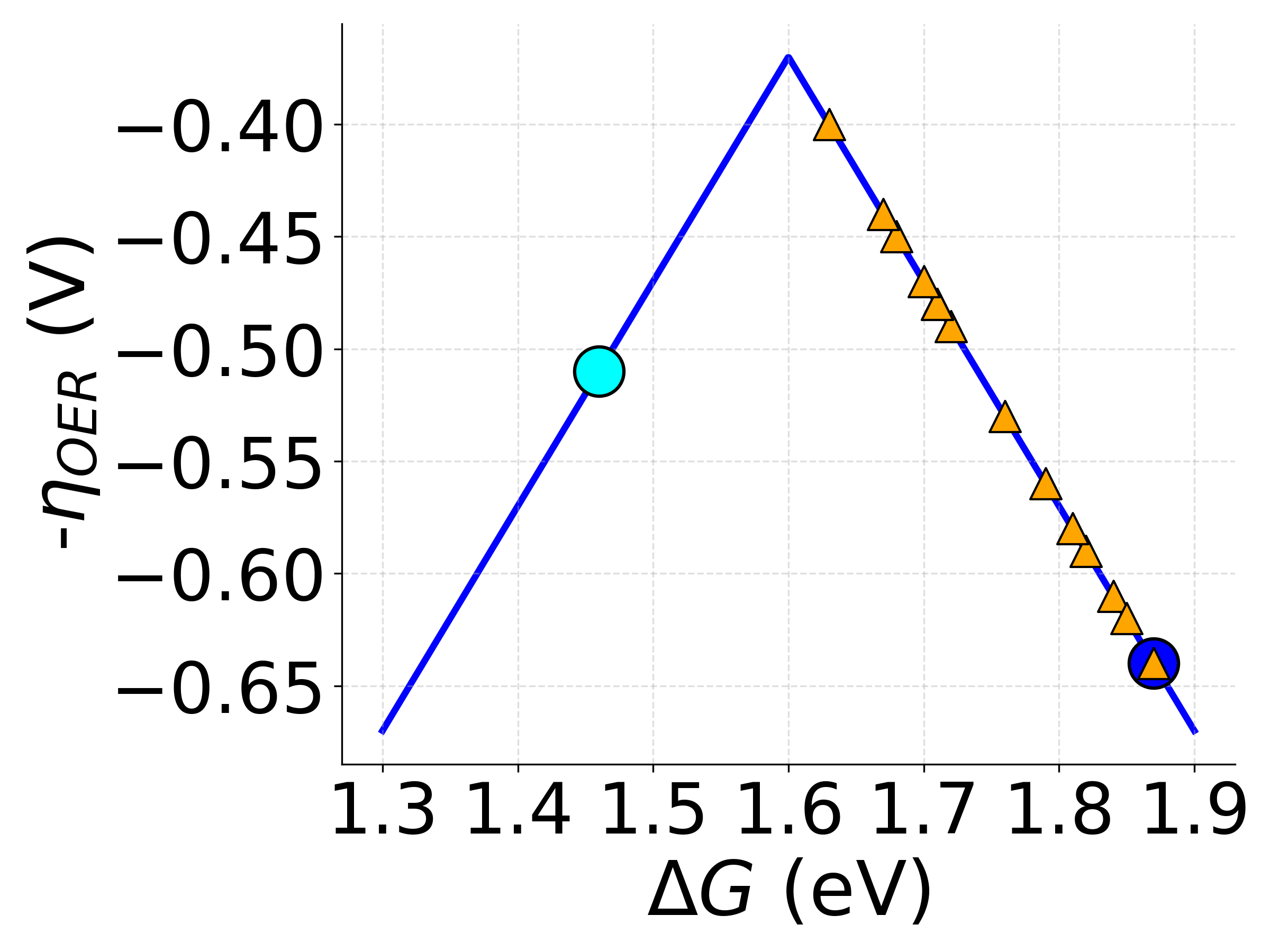}\\(b)}\\[1ex]
  \parbox{1.7in}{\includegraphics[width=1.7in]{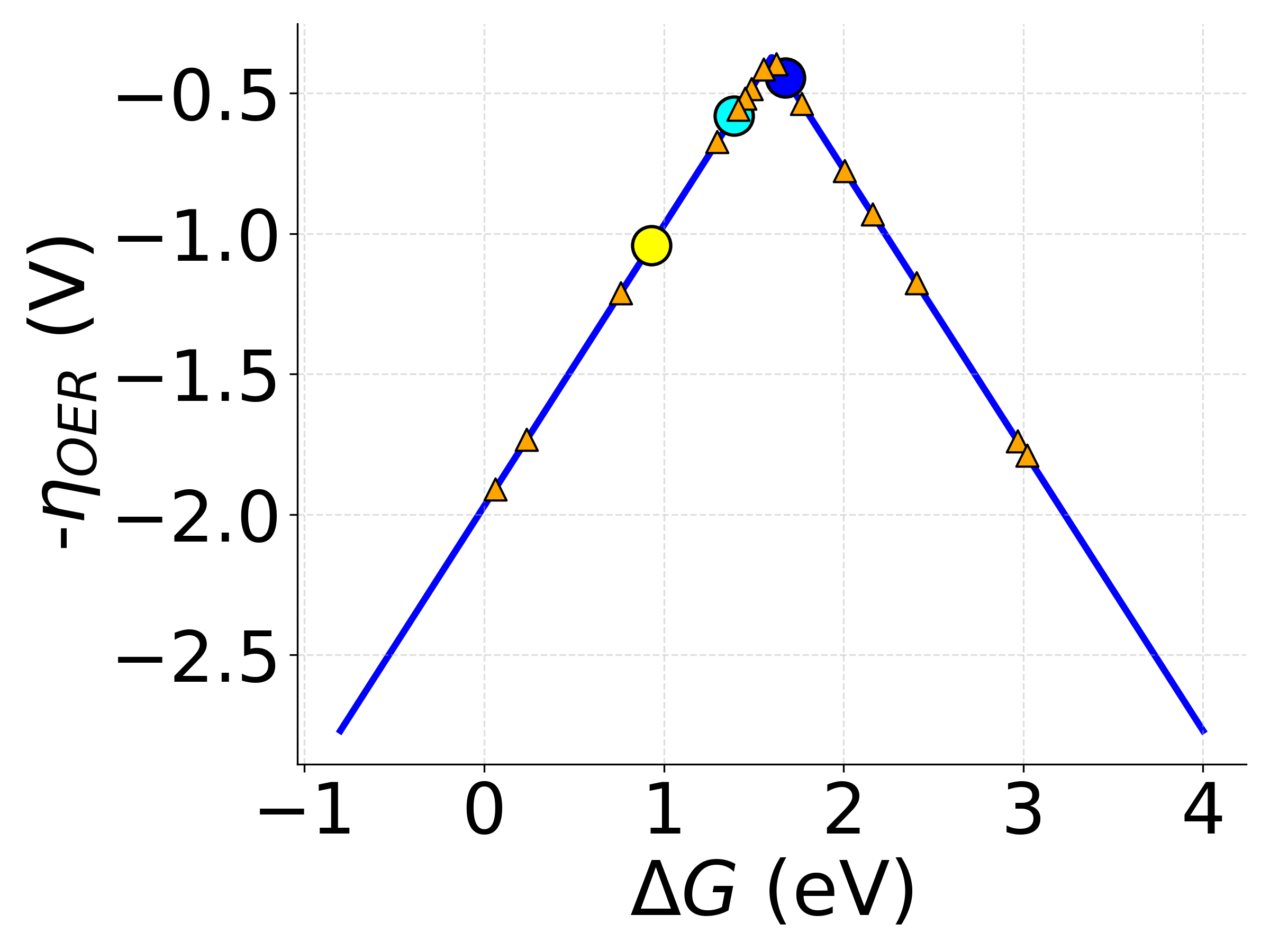}\\(c)}\hfill
  \parbox{1.7in}{\includegraphics[width=1.7in]{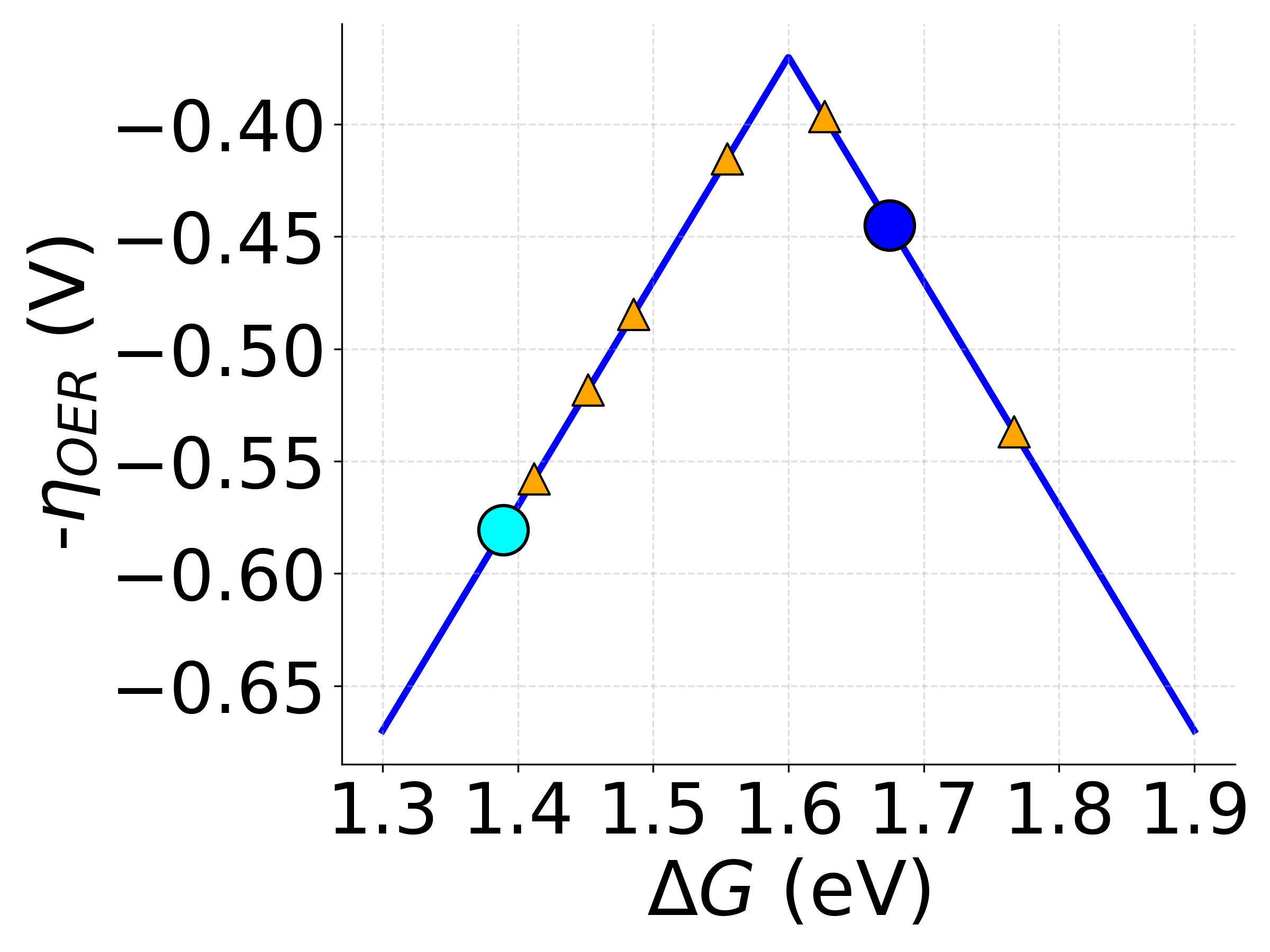}\\(d)}
  \caption{(a,b) Predicted nonmagnetic $\eta_{\rm OER}$ for pure CoO (cyan), CrO (green), NiO (blue), VO (yellow) and the 15 syntropized high entropy rock salt oxides found using EAT (triangles); (b) is a zoom of (a) near the volcano peak. (c,d) Predicted $\eta_{\rm OER}$ after enabling spin-polarization, Hubbard-$U$, and relaxing the slab. CrO has outlying $\Delta G = -5.44\,\mathrm{eV}$ and is omitted for clarity. (d) is a zoom of (c) near the volcano peak.}
  \label{fig:volcano}
\end{figure}

With $16$ effective atoms---each consisting of $4$ species---the combinatorial design space exceeds $10^9$ possible unit cells. Remarkably, EAT typically locates a near‐optimal cell in $\sim50$ iterations, each requiring only two DFT evaluations (one for O$^*$ and one for HO$^*$). This represents an order of magntitude fewer DFT evalations than involved in a typical genetic or Bayesian search. Figs.~\ref{fig:volcano}(a,b) show that all but one EAT prediction improves upon pure NiO and that many lie close to the volcano peak (triangles, Figure~\ref{fig:volcano}(a,b)), with the leading candidate reaching $\Delta G = 1.63$~eV, within 0.03~eV of the target we set.

To finalize our search, we next include spin polarization, Hubbard-$U$, and full ionic relaxation effects. We begin with the above 15 EAT‐optimized compositions and perform final DFT calculations while accounting for these last effects. Figs.~\ref{fig:volcano}(c,d) show the resulting magnetic, ion‐relaxed overpotentials. The single‐metal oxides shift markedly: in particular, CrO reconstructs dramatically upon O–adsorption, yielding an outlying $\Delta G = -5.44\,\mathrm{eV}$ (omitted for clarity).

Despite these added complexities, many high‐entropy candidates still achieve near‐optimal $\Delta G$, with the best reaching $\Delta G = 1.63\,\mathrm{eV}$ (coincidentally the same numerical value as the nonmagnetic case, but now obtained after including magnetism, Hubbard-U, and structural relaxation), twice as close to the volcano peak as pure NiO. Notably, the most active surfaces are high‐entropy oxides in which Ni and V dominate but Co and Cr remain present at meaningful levels. Under uniform mixing, the probability of realizing our best configuration across the surface is maximized by adopting a bulk stoichiometry proportional to its site occupancies, namely Co$_{0.19}$Cr$_{0.06}$V$_{0.31}$Ni$_{0.44}$O, our recommended recipe/composition. We further find that this composition is energetically stable relative to competing binary oxides---often the dominant decomposition products under oxidizing conditions~\cite{hautier2012oxides}---providing evidence that this may be a viable electrocatalyst.

\textit{Application to bonding-topology optimization}---EAT can be extended beyond fixed-topology chemical ordering because it already permits full relaxation of local atomic coordinates and lattice degrees of freedom. To allow the system to escape one bonding topology and reach another, we introduce an additional species: a ``null'' element type. Real elements mix continuously with this null type, allowing atoms to smoothly disappear from one location and reappear at another without traversing the intervening high-energy configurations along the physical path. This allows the optimization to move continuously between bonding topologies within the EAT framework. For this first implementation, we combine gradient relaxation with a simple, greedy exploratory ``shock'' scheme to help escape local minima (details are provided in the End Matter).

\begin{figure}[tb]
\centering
\parbox[c]{0.6\columnwidth}{%
    \centering
    \includegraphics[width=0.60\columnwidth]{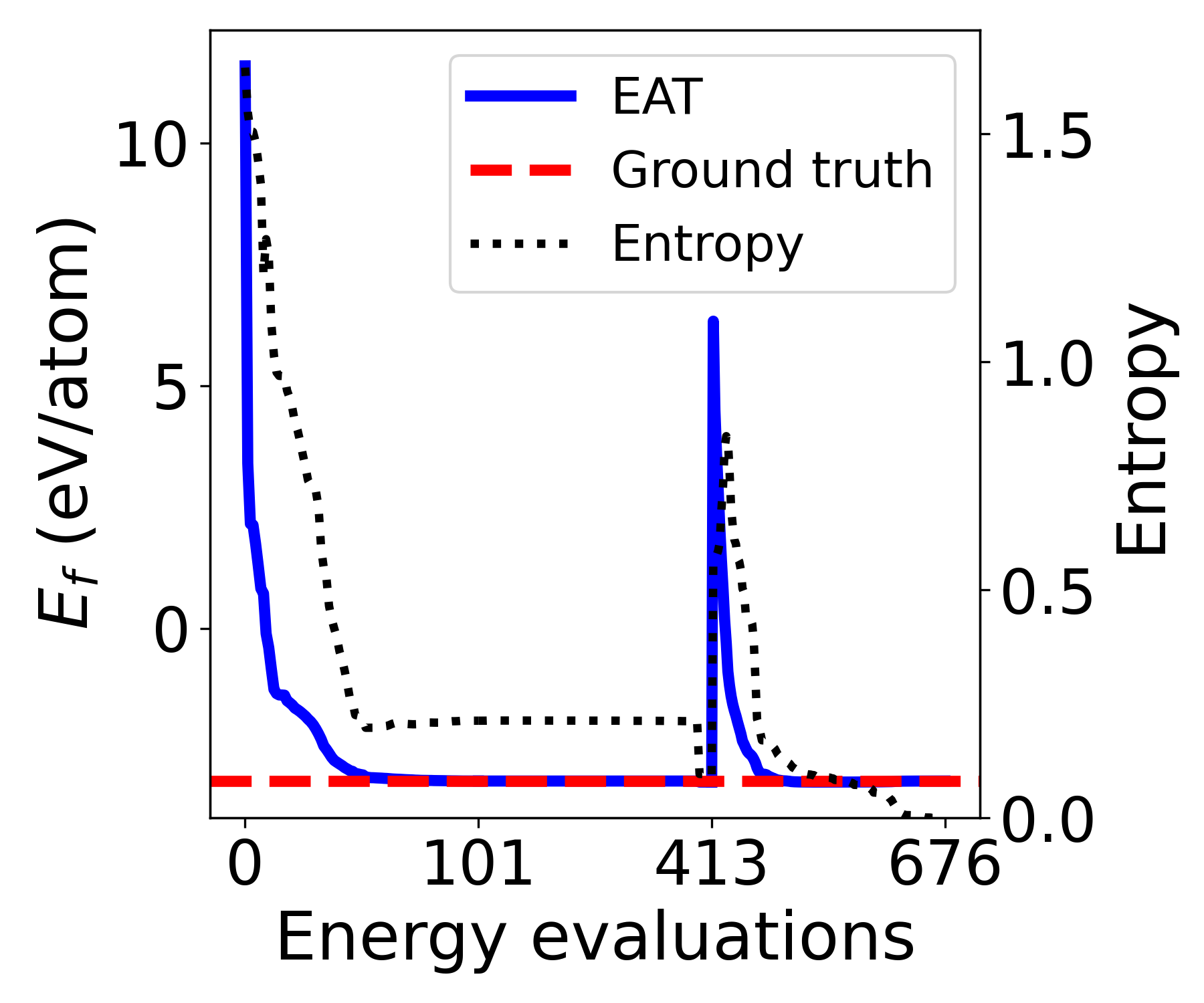}\\
    (a)
}
\hfill
\parbox[c]{0.3\columnwidth}{%
    \centering
    \includegraphics[width=0.22\columnwidth]{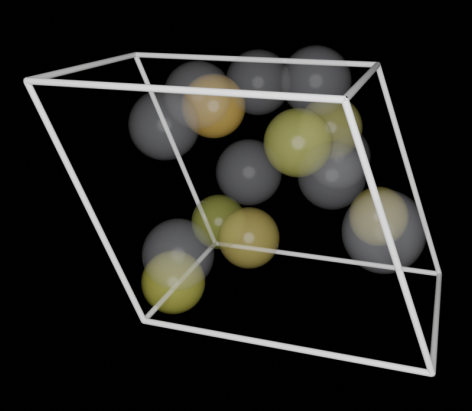}\\
    (b)
    
    \vspace{0.5em}
    
    \includegraphics[width=0.22\columnwidth]{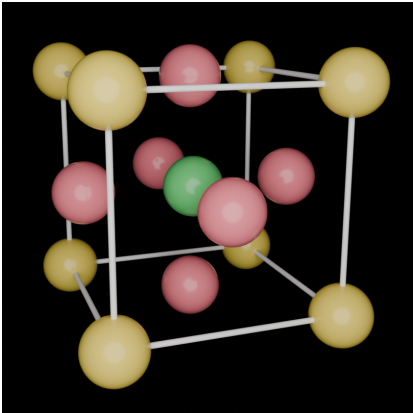}\\
    (c)
}
\caption{
Simultaneous EAT optimization of chemical ordering and structural topology for SrTiO$_3$.
(a) Energy (blue curve) and species entropy (dotted black curve) versus iteration, with ground-truth energy of SrTiO$_3$ indicated by the red dashed line. Randomizing shock events occur at iterations 101 and 413, with the first rejected and the second accepted.
(b) Initial random configuration showing lattice vectors, atom locations, and species mixtures.
(c) Final syntropized structure showing the optimized lattice vectors, bonding topology and chemical ordering. In (b,c) atom colors show corresponding mixture of Sr (yellow), Ti (green), O (red), and null (transparency), with periodic images shown for atoms which lie on the boundaries of the unit cell.
}
\label{fig:eat_topology}
\end{figure}

Figure~\ref{fig:eat_topology} shows topology optimization using the FAIRChem implementation of EAT for SrTiO$_3$, starting from random lattice vectors, ionic positions, and species weights for a cell with 16 effective atoms, with no information about the final atom count beyond the target relative elemental abundances. In Fig.~\ref{fig:eat_topology}(a), the energy initially reaches a value near the ground state, while the entropy remains nonzero---indicating that probabilistic effective atoms remain. Following a later shock and relaxation, the energy again approaches this value while the entropy falls to zero, yielding a structure consisting of atoms of definite chemical identity arranged in the correct SrTiO$_3$ perovskite bonding topology, with its characteristic corner-sharing TiO$_6$ network and Sr atoms correctly occupying the perovskite $A$-sites.   Figures~\ref{fig:eat_topology}(b,c) show, respectively, the random initial configuration and the final optimized structure reached after only 700 energy evaluations. The low-temperature ten-atom antiferrodistortive ground state is only 1 meV/atom lower in energy in the FAIRChem potential than the well-known five-atom room-temperature cubic phase recovered here, a difference that approaches the numerical noise level of the model and that may help to explain why the current implementation does not find the lower-energy state. Although still preliminary, this bonding-topology version of EAT is competitive with early discrete evolutionary searches, which required hundreds of relaxed candidates---and thus many thousands of energy evaluations~\cite{glass2006uspex}. We expect the performance of null-atom EAT for topology optimization to improve further as the method matures, especially through the combination of EAT-based continuous relaxation with population-level genetic operations.


\textit{Conclusion}---We have presented Effective Atom Theory (EAT), a route to materials optimization that embeds discrete atomic identities in a smooth, differentiable space and thereby transforms combinatorial materials search into gradient-driven optimization. Across examples spanning ground-state chemical ordering, Cu–Pd phase-diagram construction, DFT-based oxygen-evolution catalyst optimization, and SrTiO$_3$ bonding-topology optimization, EAT yields fully discrete, physically realizable structures while requiring orders of magnitude fewer evaluations than direct discrete search. Because the method can optimize any property expressible through total-energy differences—including adsorption energies, defect formation energies, mixing enthalpies, and electronic descriptors—it provides a general framework for on-the-fly discovery of multicomponent materials. By combining continuous stoichiometry variables, syntropization, and analytic gradients, EAT opens a practical route through the combinatorial wall that has long limited first-principles materials design.

\textit{Acknowledgments}--- This work was primarily supported by the Center for Alkaline-Based Energy Solutions (CABES), part of the Energy Frontier Research Center (EFRC) program supported by the U.S. Department of Energy, under grant DE-SC-0019445. C.M. was supported by from the US National Science Foundation (NSF) under award PHY-1549132, the Center for Bright Beams (CBB).

\textit{Data availability}--- The data and code that support the findings of this Letter are openly available~\cite{Tahmassebpur2025EATdata}.

\bibliographystyle{apsrev4-2}
\bibliography{main}

\clearpage

\section{End Matter}

\textit{DFT with effective atoms---}We consider $N_{\rm ion}$ effective atoms at positions $\bm R_I$ and $N_{\rm el}$ electrons with Kohn–Sham orbitals $\psi_i$ and occupations $f_i$.  Introducing the atomic stoichiometries $x_{I\alpha}$ and the effective nuclear charge 
\[
  \tilde Z_I \;=\;\sum_\alpha x_{I\alpha}\,Z_\alpha,
\]
with $Z_\alpha$ being the nuclear charge of species $\alpha$, the DFT ground‐state energy (already minimized over $\{\psi_i,f_i,\bm R_I\}$) becomes
\begin{align}
    E[\{x_{I\alpha}\}] 
    =& T[\{\psi_i, f_i\}] + E_{H,xc}[\{\psi_i, f_i\}] \notag \\ 
     & +\sum_{i,I,\alpha} f_i \bra{\psi_i}x_{I\alpha}\hat{V}_{\alpha}(\boldsymbol{R}_I)\ket{\psi_i} \notag \\
     & +\frac{1}{2}\sum_{I,J\neq I}\frac{\tilde{Z}_I \tilde{Z}_J}{|\boldsymbol{R}_I - \boldsymbol{R}_J|},
    \label{eq:EAT_Energy}
\end{align}
where $T$ is the non-interacting free energy including electron kinetic energy and electronic entropy, $E_{H,xc}$ the Hartree+XC term, and $\hat V_\alpha$ the local+nonlocal potential for the elemental species $\alpha$. Note that in Eq.~\eqref{eq:EAT_Energy}, the pseudopotentials of the effective atoms are treated as a linear combination of elemental pseudopotential operators $\hat{V}_\alpha$, weighted by $x_{I\alpha}$. Ordinary ions follow by setting $x_{I\alpha}=\delta_{\alpha,s_I}$ where $s_I$ is the species of ion $I$. This construction is similar to the VCA in \cite{Bellaiche2000VCA} and its implementation in the Vienna ab initio Simulation Package (VASP).

For $\psi_i,f_i,\bm R_I$ satisfying the stationary conditions (fully relaxed wave functions, occupancies, and all ionic positions relaxed or held fixed), the total derivative with respect to\ $x_{I\alpha}$ can be evaluated with the Hellmann-Feynman theorem, yielding 
\begin{align}
    \frac{dE(\{x_{I\alpha}\})}{dx_{I\alpha}} 
    =& \mu Z_\alpha + \sum_i f_i \bra{\psi_i}\hat{V}_{\alpha}(\boldsymbol{R}_I)\ket{\psi_i} \notag \\ 
    &+\sum_{J\neq I}\frac{Z_\alpha \tilde{Z}_J}{|\boldsymbol{R}_I - \boldsymbol{R}_J|},
    \label{eq:EAT_gradient}
\end{align}
where $\mu = \partial E / \partial N_{\rm el}$ is the electron chemical potential, representing the change in energy due to the variation in the number of electrons associated with changing the number of atoms of species~$I$. Note that, importantly, as long as the ionic positions are fully relaxed (or the ions are held frozen), no \emph{additional} self-consistent-field (SCF) cycles, terms associated with motion of the ions, or atomic relaxations are required to evaluate these derivatives. We therefore have access to these derivatives at essentially zero additional computational cost. 

\textit{FAIRChem with effective atoms}---In FAIRChem, an atomic structure is encoded as a graph neural network: atoms are graph nodes, and edges connect nearby atoms. To incorporate effective atoms, we replace the discrete elemental identity at each site by a weighted superposition of elemental representations. Concretely, each chemical species $\alpha$ has a learned node (``sphere'') embedding $h_\alpha$, as well as learned edgepoint embeddings for its roles as an edge source and edge target, denoted $g^{\mathrm{s}}_\alpha$ and $g^{\mathrm{t}}_\alpha$, respectively~\cite{wood2025uma}. If site $I$ is an effective atom with mixing coefficients $x_{I\alpha}$, then its node and edge representations are obtained by linear interpolation over species:
\begin{align}
h_{\mathrm{eff},I} &= \sum_\alpha x_{I\alpha}\, h_\alpha, \\
g^{\mathrm{s}}_{I} &= \sum_\alpha x_{I\alpha}\, g^{\mathrm{s}}_\alpha, \qquad
g^{\mathrm{t}}_{I} = \sum_\alpha x_{I\alpha}\, g^{\mathrm{t}}_\alpha .
\end{align}
To maintain consistency with FAIRChem's energy calibration, we apply the same linear mixing to the per-element reference energies $e_\alpha$~\cite{wood2025uma}, giving
\begin{equation}
e_I = \sum_\alpha x_{I\alpha}\, e_\alpha .
\end{equation}
To enable bonding topology optimization, the mixing coefficients include an additional null component, $x_{I\emptyset}$. Per-site energy contributions are multiplied by $x_{I\emptyset}$, and messages along an edge $(I,J)$ are multiplied by $x_{I\emptyset}x_{J\emptyset}$.

\textit{DFT details}---For our high entropy oxide DFT calculations, we use the open‐source plane‐wave code \texttt{JDFTx}~\cite{jdftx}, with norm‐conserving pseudopotentials~\cite{schlipf2015}, the RPBE exchange functional~\cite{norskov_RPBE}, and the PBE correlation functional~\cite{PBE_functional}. We sample the Brillouin zone with a $3\times3\times1$ $\mathbf{k}$‐point grid, employ a $30$~H (Hartree) wavefunction cutoff and a $120$~H density cutoff, and apply Fermi–Dirac smearing at an effective temperature of $315$~K for the Kohn–Sham occupancies. We converge electronic energies to $10^{-5}$~H and employ a $\sqrt{2}\times\sqrt{2}\times2$ rock‐salt (100) facet slab containing 16 metal and 16 oxygen atoms, with a $4.3$~\AA\ lattice constant and $13$~\AA\ of vacuum. We seed spin‐polarized DFT calculations with NiO’s AFM configuration~\cite{AFM_NiO} and apply Hubbard–$U$ values of 3.3~eV (Co), 3.5~eV (Cr), 3.1~eV (V), and 6.4~eV (Ni)~\cite{hubbU}. The adsorbate and top two slab layers are fully relaxed for these magnetic calculations.

For our Cu--Pd convex hull DFT calculations, we use the open-source plane-wave code \texttt{JDFTx}~\cite{jdftx}, with norm-conserving pseudopotentials~\cite{schlipf2015} and the PBE exchange--correlation functional~\cite{PBE_functional}. We employ a $30$~H wavefunction cutoff and a $120$~H density cutoff, apply Fermi--Dirac smearing at an effective temperature of $630$~K for the Kohn--Sham occupancies, and converge electronic energies to $10^{-6}$~H. For the $1\times2\times2$ conventional FCC alloy cell, the Brillouin zone is sampled using an $8\times4\times4$ $\mathbf{k}$-point grid, while for the $1\times1\times1$ conventional FCC reference cells we use an $8\times8\times8$ $\mathbf{k}$-point grid.

\textit{Syntropization details}---In practice, syntropization is controlled by four hyperparameters: the initial penalty strength $\lambda_0$, a multiplicative ramp factor $\tau$, the initialization distribution for the atom stoichiometries $x_{I\alpha}$, and the Tsallis-$q$ parameter. We benchmark these choices in the Supplementary Material. We start with a small $\lambda_0$ (e.g., $10^{-3}$) so that the early iterations prioritize minimizing the target objective rather than the entropy penalty. After the optimization converges for the initial fixed $\lambda=\lambda_0$ (no $\tau$ present yet), we then modify the penalty by $\lambda \leftarrow \tau\lambda$, warm-start from the previous solution, and re-optimize; we repeat this until the $x_{I\alpha}$ all become ones and zeros. With testing, we find that $\tau=1.2$ and $q=2$ work well. For initialization, we find that optimal performance is achieved when the $x_{I\alpha}$ are drawn from a symmetric Dirichlet distribution with inverse variance parameter $\alpha_{\mathrm{Dir}}=5$ \cite{scipy}. Since for $\alpha_{\mathrm{Dir}}\rightarrow 1$ the distribution becomes uniform, and for $\alpha_{\mathrm{Dir}}\to\infty$ the distribution becomes a Dirac delta distribution at the center of the simplex, our initialization scheme thus avoids being overly concentrated near the simplex center or overly uniform over the simplex. (See the Supplemental Material for detailed test data.)

\textit{Mixing elements with differing valence orbitals}---Within DFT, EAT works particularly well for combinations of elements with reasonably compatible valence electronic structures. In practice, creating effective atoms from species with incompatible valence electronic structures (i.e. different occupied angular‐momentum channels)---for example, alkali and transition metals, or transition metals and main‐group nonmetals---can yield DFT energy landscapes that are continuous but with rapid variations in gradients that are numerically challenging to optimize without more advanced techniques. Note that forming, for example, mixtures of alkali elements on some sites and mixtures of transition metals on other distinct sites poses no issue. The difficulty arises only when combining species with incompatible valence structures within a single effective atom. We note that the FAIRChem implementation of EAT does not share this issue, and so any combination of elements may be constructed.

 \textit{Details for NaCl ground-state demonstration}---
 The objective is to minimize the formation energy subject to a fixed global composition of 50/50 Na and Cl, which is imposed by a composition penalty on the objective function,
\begin{equation}
\frac{\xi N}{2}\left(\frac{1}{N}\sum_{I=1}^{N} x_{I,\mathrm{Na}} - t\right)^2,
\label{eq:composition}
\end{equation}
where $N$ is the number of sites in the supercell, $\xi=1$ is a constant determining the strength of the penalty, and $t=0.50$. In FAIRChem, the total formation energy is
\begin{equation}
E_f(\boldsymbol{x})
=
E_{\mathrm{alloy}}(\boldsymbol{x})
-
\sum_{i,\alpha} x_{I\alpha} E_{\mathrm{ref},\alpha},
\label{eq:Ef_FAIRChem}
\end{equation}
where $E_{\mathrm{ref},\alpha}$ is a fixed reference energy for species $\alpha$.
 For the ground-state chemical ordering application, to summarize EAT's performance we consider random subsets $D_n$ of size $n$ drawn without replacement from the 100 runs, and define
\begin{equation}
M_n \;=\; \min_{r\in D_n} E_{f,r}^{\mathrm{final}},
\end{equation}
where $E_{f,r}^{\mathrm{final}}$ is the final (syntropized) formation energy returned by run $r$ at the target composition.
We report the median of $M_n$ over many random subset draws and plot it against an \emph{effective} evaluation count $nL$, which accounts for the fact that each EAT run costs $\approx L$ energy evaluations. Under this metric, when the median reaches the ground-truth formation energy, EAT has a 50\% probability of having found the ground state using $nL$ energy evaluations.

As a baseline, we compare EAT with sequential Gaussian-process (GP) Bayesian optimization in BoTorch over fixed-composition atomic orderings \cite{balandat2020botorch}. For this comparison, each structure is encoded by categorical site-occupation variables, with the target composition enforced exactly by a repair operator. We fit the GP surrogate to the negative formation energy and select new candidates sequentially using analytic Log Expected Improvement over a discrete candidate pool. At each step, this pool combines random composition-preserving orderings with local proposals generated by random species swaps in the current best structure.

\textit{Formation energy in DFT}---When using effective atoms in DFT, the \textit{total} formation energy is
\begin{equation}
E_f(\boldsymbol{x})
=
E_{\mathrm{alloy}}(\boldsymbol{x})
-
\sum_{I=1}^{N} E_{\mathrm{ref}}(\boldsymbol{x}_{I}),
\label{eq:Ef_DFT}
\end{equation}
where $E_{\mathrm{alloy}}(\boldsymbol{x})$ is the energy per atom of the alloy supercell and
$E_{\mathrm{ref}}(\boldsymbol{x}_{I,\cdot})$ is the energy per atom of the reference material corresponding to the effective-atom stoichiometry on site $I$.

\textit{Optimizing Structural Topology with EAT}---We begin from a random unit cell containing a specified number of randomly placed effective atoms with random species weights. The optimization then proceeds as in standard EAT syntropization, except that we periodically apply greedy ``shocks'' to encourage escape from local minima. These shocks are triggered when the combined objective function, excluding the species-entropy term, has plateaued to within 10$^{-3}$\,eV for 3 iterations. For each shock event, we run 3 trial perturbations: we apply random Gaussian perturbations of $\sigma$=0.25\,\AA\ to the positions and $\sigma$=0.35 to the species weights of all non-null effective atoms, apply random Gaussian perturbations of length $\sigma$=0.25\,\AA\  and angles $\sigma$=4.0\,degrees to the unit cell, and move all fully null effective atoms to new uniform random positions within the cell.  Each shocked configuration is then relaxed, and the trial is accepted only if the resulting combined objective function, again excluding species entropy, is lower than its value before the shock.



\end{document}